\documentclass[journal,twoside,web]{IEEEtran}
\usepackage{tmi}
\usepackage{cite}
\usepackage{amsmath,amssymb,amsfonts}
\usepackage{textcomp}

\usepackage{booktabs}
\usepackage{float}
\usepackage{multicol}
\usepackage{multirow}
\usepackage{stfloats}
\usepackage{graphicx}
\usepackage{algorithm}
\usepackage{algpseudocode}
\def\BibTeX{{\rm B\kern-.05em{\sc i\kern-.025em b}\kern-.08em
    T\kern-.1667em\lower.7ex\hbox{E}\kern-.125emX}}
\begin{document}
\title{Interactive Segmentation for COVID-19 Infection Quantification on Longitudinal CT scans}
\author{Michelle Xiao-Lin Foo, Seong Tae Kim, Magdalini Paschali, Leili Goli, Egon Burian, Marcus Makowski, Rickmer Braren, Nassir Navab, Thomas Wendler
\thanks{
This work was partially supported by the Bavarian Research Foundation (BFS, grant AZ-1429-20), the Bavarian State Ministry for Economics, Development and Energy (BayStMW, DIK 0127/02), the Korean Government (MSIT) under Institute of Information and Communications Technology Planning and Evaluation (IITP) Grant 2021-0-02068 and National Research Foundation of Korea (NRF) Grant 2021R1G1A1094990. (Corresponding author: Seong Tae Kim)}
\thanks{M.X.-L. Foo, M. Paschali, N. Navab, and T. Wendler are with Chair for Computer Aided Medical Procedures and Augmented Reality, Technical University of Munich, Germany.}
\thanks{S.T. Kim is with the Department of Computer Science and Engineering, Kyung Hee University, South Korea (email: st.kim@khu.ac.kr)}
\thanks{L. Goli is with Department of Electrical and Computer Engineering, University of Toronto, Canada}
\thanks{E. Burian, M. Makowski, and R. Braren are with Institute for Diagnostic and Interventional Neuroradiology, Technical University of Munich, Germany}
}

\maketitle

\begin{abstract}
Consistent segmentation of COVID-19 patients' CT scans across multiple time points is essential to assess disease progression and response to therapy accurately. Existing automatic and interactive segmentation models for medical images only use data from a single time point (static). However, valuable segmentation information from previous time points is often not used to aid the segmentation of a patient's follow-up scans. Also, fully automatic segmentation techniques frequently produce results that would need further editing for clinical use. 
In this work, we propose a new single network model for interactive segmentation that fully utilizes all available past information to refine the segmentation of follow-up scans. In the first segmentation round, our model takes 3D volumes of medical images from two-time points (target and reference) as concatenated slices with the additional reference time point segmentation as a guide to segment the target scan. 
In subsequent segmentation refinement rounds, user feedback in the form of scribbles that correct the segmentation and the target’s previous segmentation results are additionally fed into the model. This ensures that the segmentation information from previous refinement rounds is retained. Experimental results on our in-house multiclass longitudinal COVID-19 dataset show that the proposed model outperforms its static version and can assist in localizing COVID-19 infections in patient's follow-up scans. 
\end{abstract}

\begin{IEEEkeywords}
COVID-19, CT, Interactive Segmentation, Deep Learning
\end{IEEEkeywords}

\section{Introduction}
\label{sec:introduction}
\IEEEPARstart{O}{ver} the last few years, there has been immense progress in the medical image analysis field due to the introduction of deep learning networks. Specifically, such automatic segmentation methods are nowadays widely adopted in data annotation software. 

In December 2019, the first cases of a new coronavirus disease, COVID-19, a severe acute respiratory illness, emerged in Wuhan, China~\cite{zhou2020ct}. This highly infectious respiratory virus rapidly spread worldwide and threw the world into a global pandemic. According to the numbers on the COVID-19 monitoring site by Johns Hopkins University\footnote{\texttt{https://coronavirus.jhu.edu/map.html}}, as of August 30th, 2021, more than 216 million people have been infected and 4.498 million people have succumbed due to the complications caused by the virus.

During the initial outbreak of COVID-19, there was an urgent need for fast annotation of medical scans in order to further understand the disease. Numerous researchers utilized deep learning-based methods for this task~\cite{Wang2020,Shan2021,Fan2020}. Computed tomography (CT) scans provide crucial diagnostic information in the assessment and treatment of COVID-19 patients~\cite{Ai2020,Li2020}. However, medical reports show observations that the imaging features of COVID-19 are mixed and diverse among patients, and the change in radiological patterns during the course of the disease is inconsistent~\cite{Zhou2020,Shi2020,Mendel2020}. The subtle anatomical boundaries and variations in size, density, location, and texture of the disease pose a challenge for automatic segmentation techniques.

Human interactions coupled with deep learning models can offer a way to overcome the challenges faced by automatic segmentation models and to improve segmentation results as shown by~\cite{Ramadan2020,Wang2018,DBLP:conf/miccai/ZhouCW19}. However, previous works on interactive segmentation only used single time point data for segmentation. The readily available segmentation information from previous time points has not been exploited to segment a patient's follow-up scans. 
  
To address the limitations of automatic static segmentation models, in this paper, we propose an interactive segmentation method that segments COVID-19 infection on longitudinal CT scans. The proposed method is designed to leverage available longitudinal information and user feedback to improve segmentation quality.   
The main contributions of this paper can be summarized as follows: 
\begin{itemize}
    \item We propose a new segmentation approach that utilizes information from previous time point, past segmentation refinement rounds, and user feedback. To the best of our knowledge, it is the first work to design an interactive segmentation on longitudinal CT scans. Our method is also the first interactive segmentation of COVID-19 infection using longitudinal COVID-19 CT scans.
    \item Our method can be used to extend existing static models for longitudinal interactive segmentation with minimal effort. 
    \item We conduct an extensive study using an in-house longitudinal COVID-19 dataset to showcase the improved performance of the longitudinal interactive segmentation model over the static interactive segmentation model.
\end{itemize}

\section{Related Works}

\subsection{COVID-19 Infection CT Segmentation}
After the outbreak of COVID-19, in order to understand the disease and support the radiologists several studies were presented. 
For the task of learning from noisy labels to segment COVID-19 pneumonia lesions from lung CT scans, Wang et al.~\cite{Wang2020} proposed a noise-robust training method. Shan et al.~\cite{Shan2021} presented a modified 3D convolutional neural network called VB-Net, which combines V-Net~\cite{2016arXiv160604797M}, a fully convolutional neural network for volumetric medical image segmentation with a bottle-neck deep residual learning framework for quantitative COVID-19 infection assessment. Fan et al.~\cite{Fan2020} proposed a semi-supervised learning approach for segmenting diverse radiological patterns such as ground-glass opacity and consolidation from lung CT scans. However, due to the subtle anatomical boundaries, pleural-based location, and high variations in infection characteristics, it is still challenging to automatically identify and quantify CT image findings related to COVID-19~\cite{Zhou2020,Shi2020}.

\subsection{Interactive Segmentation}
In interactive segmentation, feedback from users is used to enhance the predictions of machine learning models. This human-in-the-loop method shows potential in improving segmentation results~\cite{Ramadan2020}. However, interactive segmentation needs to be accurate and efficient in order to be helpful in a clinical setting. Xu et al.~\cite{Xu2016} introduced a deep interactive object selection method where user-provided clicks are transformed into Euclidean distance maps. However, Euclidean distance does not exploit image context information. In contrast, the geodesic distance transform by \cite{Criminisi2008} further encodes spatial regularization and contrast-sensitivity information but suffers from sensitivity to noise. The deep learning-based interactive segmentation framework by Wang et al.~\cite{Wang2018} incorporated user-provided bounding boxes and scribbles (lines drawn on wrongly segmented areas). With the inclusion of this user feedback, they demonstrated an increase in segmentation performance. Zhou et al.~\cite{DBLP:conf/miccai/ZhouCW19} showed that with a small number of user interactions, segmentation accuracy can be substantially improved. Kitrungrotsakul et al.~\cite{kitrungrotsakul2021attention} proposed a segmentation refinement module that can be appended to automatic segmentation networks and utilized a skip connection attention module to improve important features for both segmentation and refinement tasks. Withal, the methods introduced above are designed for single time point data.

\subsection{Longitudinal Image Segmentation}
To effectively study how a disease progresses, consistent segmentation of the affected regions based on scans from multiple time points can further provide important information. Birenbaum et al.~\cite{birenbaum} suggest the use of multiple longitudinal networks to process longitudinal patches from different views where the model concatenates the output of the encoder to produce Multiple Sclerosis (MS) lesion segmentation. Their method shows that the inclusion of information from multiple time points is beneficial to the model. To improve segmentation of MS lesions by taking advantage of the longitudinal information, \cite{Denner2020} proposes a longitudinal network with an early fusion of two-time points scans to encode the structural differences implicitly. But due to the minute structural differences in MS lesions across different time points, it is still a challenge for the network to achieve a high accuracy. Kim et al.~\cite{kim2021} present a framework that leverages spatio-temporal cues between longitudinal scans to improve quantitative assessment of the progression of COVID-19 infection in chest CT scans. However, the implementations above do not utilize the available previous time scan segmentation mask to segment a patient's follow-up scans.

\section{Proposed Method}
\subsection{Interactive Segmentation Network}
The proposed method extends the baseline longitudinal network by Denner et al.~\cite{Denner2020} to fully exploit longitudinal information and user feedback for interactive segmentation. 
The baseline longitudinal model is referred to as a 2.5D model as it retains the global context of the CT volume by combining the per-slice prediction (2D) from three anatomical planes (coronal, sagittal, and axial views) to produce the segmentation for one voxel. Each slice is processed by FC-DenseNet56~\cite{jegou2017one}, which is a fully convolutional dense network for 2D segmentation.
Previous work~\cite{Roy2018,Zhang2019,Aslani2019,Alkadi2019} has shown that state-of-the-art results on various medical image segmentation problems are achievable through 2.5D approaches. 
This is due to the reasons that fully 3D approaches induce a high computational cost, and the patch-based 3D approaches lose the global structural information in the slice.
Therefore, our baseline model is built as the 2.5D model, which uses two-time points consisting of stacked 2D slices from three anatomical views. This method can preserve global information along the two axes and local information from the third axis~\cite{Zhang2019}. 

Let ${\textbf{V}}_{t+1}\in\mathbb{R}^{h\times w \times s}$ and ${\textbf{V}}_{t} \in \mathbb{R}^{h\times w \times s}$ denote the follow-up target CT volume and the reference previous volume, respectively. $h$ and $w$ are the height and width of the input, and $s$ is the number of slices in the volume. ${\textbf{X}}_{t+1}$ and ${\textbf{X}}_{t}$ indicate the individual slices of the volumes. In this study, we assume that the segmentation masks for the reference previous volume ${\textbf{S}}_t \in \{0,1\}^{C\times h\times w}$ is available. $C$ represents the number of foreground classes, and it is set to 2 in this study since we are targeting the segmentation of two foreground classes (i.e., ground-glass opacity and consolidation). In addition, let ${\textbf{E}}_{t+1} \in \{-1,1\}^{C\times h\times w }$ denote the editing masks on the target segmentation during the user feedback. Note that the segmentation masks for the reference previous volume and the editing masks are concatenated to the input. 

\subsubsection{Training}
\begin{figure}[t]
\centering
\includegraphics[width=\linewidth]{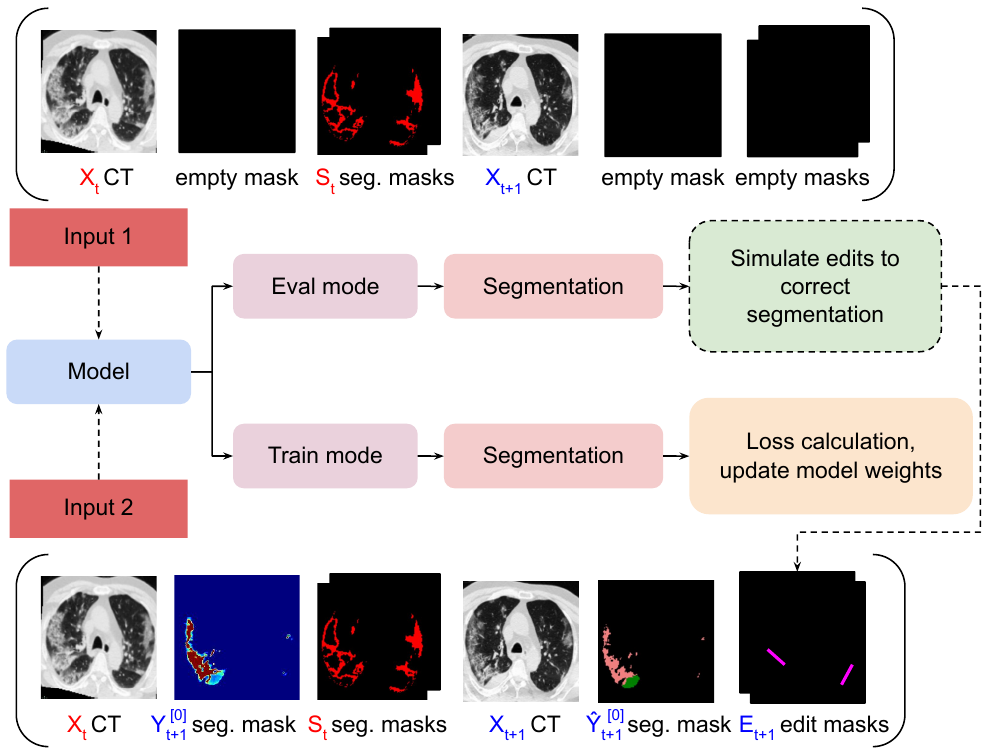}
\caption{Training flow of the proposed interactive longitudinal segmentation model. Training alternates between two training inputs that represent different scenarios: \textit{Input 1} for initial segmentation round. \textit{Input 2} for interactive segmentation rounds. Note that ${\textbf{S}}_t$ and ${\textbf{E}}_{t+1}$ two channels in our case as $C=2$. Accordingly if ${\textbf{E}}_{t+1}$ is not available as it it in the scenario of \textit{Input 1}, ${\textbf{E}}_{t+1}$ is a pair of empty masks.}
\label{fig:1}
\end{figure}

To train the model to adapt to different input information combinations during inference, throughout the training process, the model is randomly trained with two different inputs as shown in Figure~\ref{fig:1}.

\textit{Input 1} represents the scenario where the user passes the data to the model at the beginning of interactive segmentation to produce the first segmentation for the target slice, whereas \textit{input 2} represents the input data in the subsequent editing rounds. In both cases, scans from the two-time points with additional data are concatenated along the channel dimension so that structural changes that are evident between them are utilized by the model to improve its segmentation performance. Empty masks are used in place of information that is not available in the first segmentation round, such as user feedback and target prediction. 

Let $\textbf{I}^{[T]}_{t+1} \in\{0,1\}^{h\times w\times 8}$ denote the input tensor for the segmentation round $T$. The input $\textbf{I}^{[T]}_{t+1}$ consists of the reference previous CT slice $\textbf{X}_t$, the segmentation masks on the reference previous CT slice $\textbf{S}_t$, the target CT slice $\textbf{X}_{t+1}$, the highest class probability per pixel on the target slice from the previous segmentation round $\textbf{Y}^{[T-1]}_{t+1}\in \{0,1\}^{h\times w}$, the predicted class on the target slice from the previous segmentation round $\hat{\textbf{Y}}^{[T-1]}_{t+1}\in \{0,1,2\}^{h\times w}$ (0 for background, 1 for ground-glass opacity, and 2 for consolidation), the editing masks $\textbf{E}_{t+1}$ as shown in Figure 1. For \textit{Input 2}, the empty mask is used for $\textbf{Y}^{[T-1]}_{t+1}, \hat{\textbf{Y}}^{[T-1]}_{t+1}, \textbf{E}_{t+1}$.
The interactive segmentation refinement network (ISR), $f_{ISR}$ outputs the per-slice segmentation of the target image as follows:
\begin{align}
    {\textbf{Y}}^{[T]}_{t+1} = f_{ISR}(\textbf{I}^{[T]}_{t+1}) \in\{0,1\}^{h\times w}
\end{align}    
 
To produce the simulated edits for training with \textit{input 2} the model is first set to evaluation mode (without model weights updates), and an initial segmentation of the target is generated. This is then used to produce the simulated edits. The edit simulation process will be further introduced in Section~\ref{sec:is}.

During training, the segmentation of the slices is treated as a 2D segmentation problem. During inference with real user feedback, the prediction on slices from the three anatomical orientations are combined to produce the segmentation output for every voxel. The pseudocode of the training process of the proposed interactive longitudinal interactive segmentation model is presented in Algorithm \ref{alg:1}.

\begin{algorithm}
        \caption{Training of Proposed Interactive longitudinal Segmentation model, $f_{ISR}$} 
        \begin{algorithmic}[1]
                \For {$1 \leq e \leq epochs$}
                    \For {$i = 1$ to $N$}
                    \State Get 1st predictions for input batch $i$
                    \State Generate random number, $Z \in [0,1]$
                    \If{$Z > 0.5$}
                            \State Generate simulated edits for the predictions
                            \State Append simulated edits \& outputs from \\ \phantom . \phantom . \phantom . \phantom . \phantom . \phantom . \phantom .\phantom. 1st prediction round to inputs
                            \State Get 2nd predictions 
                            \State Calculate loss using 2nd predictions 
                            \Else
                            \State Calculate loss using 1st predictions  
                    \EndIf
                    \State Backpropagate \& update model weights
                    \EndFor
                \EndFor
        \end{algorithmic} 
           \label{alg:1}
\end{algorithm} 

\subsection{Edit Simulation during Training}\label{sec:is}
During training, simulated user edits for wrongly segmented regions are automatically generated. Incorrectly segmented regions are areas that are under- or over-segmented. The segmentation output from the model is compared with the ground truth to choose the slice region for simulating the user edits. Lines are automatically drawn on the selected regions as simulated feedback. As mentioned before, the edit information is concatenated to the CT scans as additional channels, one for each class, with foreground interaction having a value of 1 and background interaction -1. Because the model input is 2.5D instead of 3D, edits are simulated in the axial, coronal, and sagittal slices as opposed to only the axial slices as in~\cite{DBLP:conf/miccai/ZhouCW19}. Zhou et al.~\cite{DBLP:conf/miccai/ZhouCW19} simulate edits only for the most extensive 2D incorrectly segmented slice region. However, due to the scattered nature of the COVID-19 infections in our case, the top-5 largest wrongly segmented regions in each slice is used for edit simulation. The total number of generated edits, independent of the different classes are limited to prevent the model from overfitting and also to avoid considerable slow-down in the training process caused by the long edit simulation time needed when large numbers of incorrectly segmented regions are detected.

\begin{figure}[t]
\centering
\includegraphics[width=\linewidth]{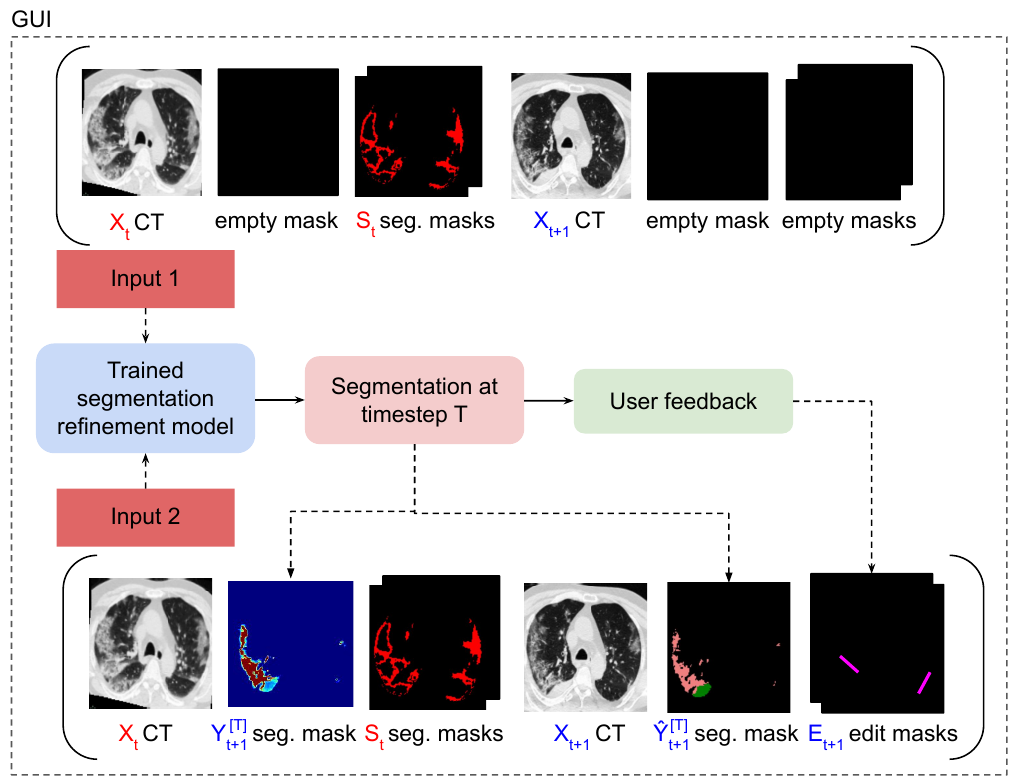}
\caption{Overview of the interactive segmentation flow with the GUI. Input to the segmentation model prior to the first user interaction resembles \textit{Input 1} that is used during training. Accordingly, input of subsequent editing rounds resembles \textit{Input 2} as in Figure~\ref{fig:1} which enables assisted segmentation refinement via the GUI.}
\label{fig:2}
\end{figure}

\begin{figure}[t]
\centering
\makebox[\linewidth][c]{\includegraphics[width=\linewidth]{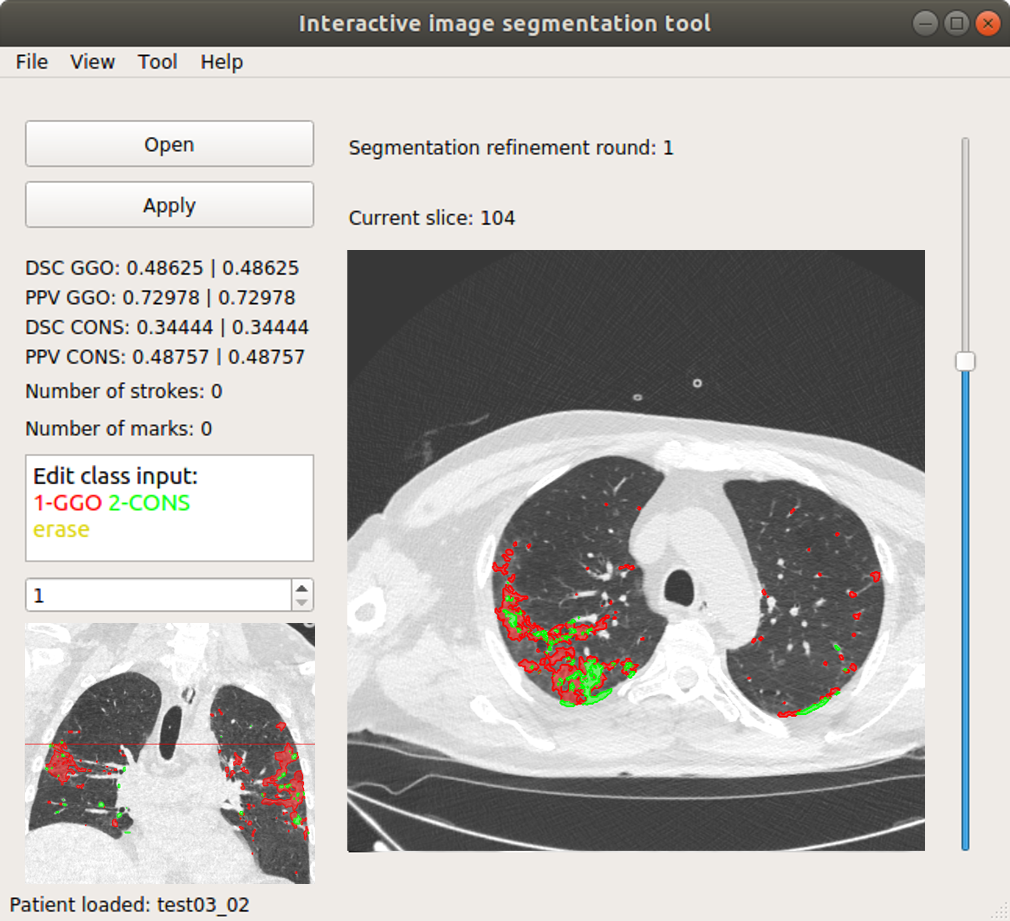}}%
\caption{Our GUI for COVID-19 lung infection interactive segmentation. The spin box in the side bar shows the class for the current brush input, here 1, i.e., the brush for ground-glass opacity (GGO).}
\label{fig:gui}
\end{figure}

\subsection{Inference with GUI for User Feedback}
Figure~\ref{fig:2} shows the interactive segmentation refinement stage. An editing graphical user interface (GUI) as shown in Figure~\ref{fig:gui} is implemented using Qt Designer\footnote{\texttt{https://doc.qt.io/qt-5/qtdesigner-manual.html}}. The GUI automatically loads the trained segmentation refinement model during start-up. Then, the user can load the CT volume to be segmented. After the initial segmentation round, the predicted segmentation will be overlaid on the scans for inspection. The user can then use the brush to edit wrongly segmented areas and run the data through the model again. This process can be repeated as many times as necessary.

For each segmentation refinement round, the user feedback on incorrectly segmented regions from current and previous rounds is summed up in each slice's edit mask. The current round user feedback has higher priority, and so it is multiplied by two before being added to the previous user edits. The values of the mask are then clipped to $[-1, 1]$. This is done so that the previous edit information is not lost. The mask is then concatenated to its corresponding slice image and fed into the segmentation refinement model.

\subsection{Implementation}
The baseline longitudinal segmentation model from~\cite{Denner2020} is modified and used in this study. It is an end-to-end 2.5D segmentation network based on FC-DenseNet56~\cite{jegou2017one} and implemented in PyTorch 1.4~\cite{NEURIPS2019_9015}. Mean Squared Error (MSE) loss, Adam optimizer with AMSGrad~\cite{47409} and a learning rate of 0.0001 are used for training. The inference time for processing a COVID-19 patient’s 2.5D data with a size of 3$\times$(150$\times$150$\times$150) takes 15 seconds on an NVIDIA GeForce RTX 2080 Ti with 11GB GPU.

\section{Experiments}
\subsection{COVID-19 Segmentation Dataset and Preprocessing}
An in-house clinical dataset collected from the Radiology Department of Technical University of Munich during the COVID-19 first wave (March-June 2020) is used for training and evaluation. It consists of 30 longitudinal low-dose native CT scans from patients age between 46 and 82 years old with a positive polymerase chain reaction (PCR) test for COVID-19. The time gap between the follow-up scans and the previous scans is 17$\pm$10 days; the scans were taken during admission and hospitalization (33$\pm$21 days, 0\textendash71 days). The scans were performed using two different CT imaging devices (IQon Spectral CT and iCT 256, Philips Healthcare, Best, the Netherlands) with the same parameters (X-ray current 140-210 mA, voltage 120kV peak, slice thickness 0.9mm) and covered the entire lung. The data was collected with the approval of the institutional review board of TUM (ethics approval 111/20 S-KH).

An expert rater (radiologist with four years of experience) annotated the dataset at voxel-level with the ImFusion Labels software (ImFusion GmbH, Munich, Germany\footnote{ImFusion, https://www.imfusion.com/}). Lung masks (lung parenchyma vs. other tissues) and pathology masks for four classes: healthy lung (HL), ground-glass opacity (GGO), consolidation (CONS), and pleural effusion (PLEFF) are generated. Due to the large variations in intensity range, size, and alignment, the raw CT volumes have to be preprocessed before they are used for training. The volumes are cropped to the lung regions using manually annotated lung masks. Intensity values outside the range (-1024, 600) are clipped, and min-max normalization is performed on the volumes before they are resized to 150$\times$150$\times$150 pixels. Slices that have a voxel-value variation smaller than 0.001\% between their min and max values are considered empty and removed. 

\begin{figure}[t]
    \centering
    \makebox[\linewidth][c]{\includegraphics[width=\linewidth]{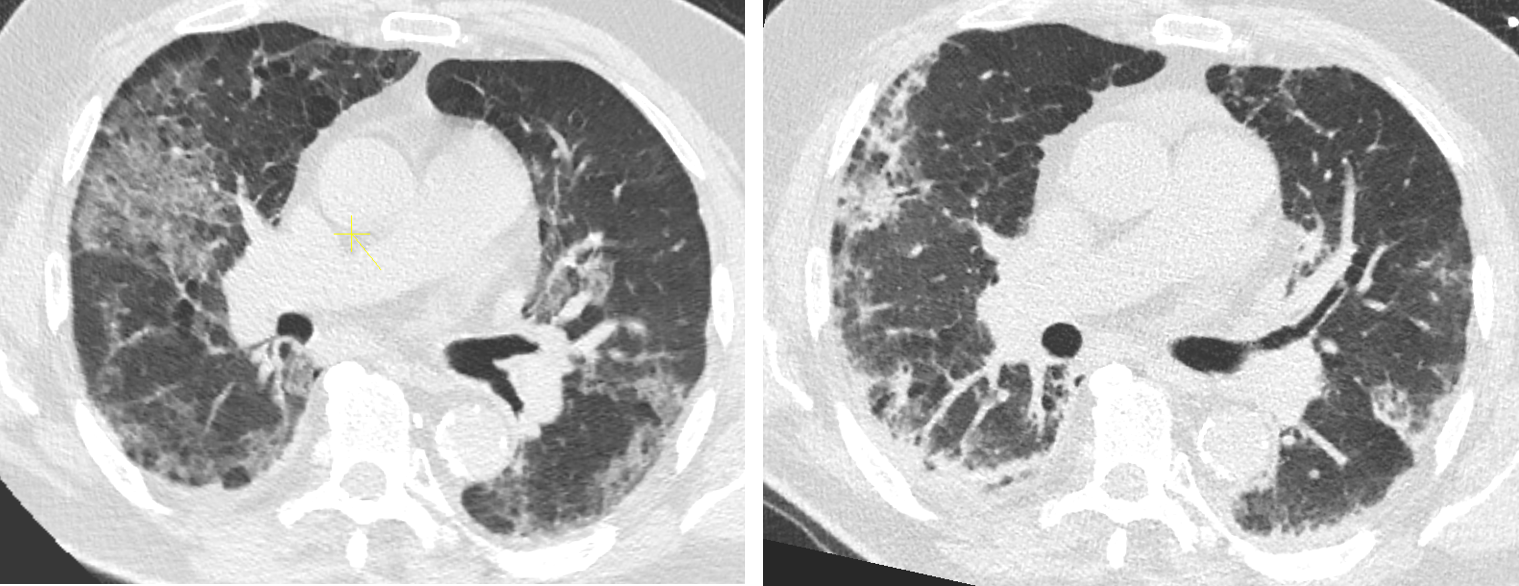}}%
    \caption{Deformable registration example of reference scan to target scan. \textit{Left:} Reference scan \textit{Right:} Target scan.}
    \label{fig:regis}
\end{figure}

Similar to \cite{kim2021}, we also use the deformable registration algorithm by \cite{Lowekamp2013}, where the image is deformed through a B-Spline Transform that uses a sparse set of grid points overlaid onto the fixed domain of the image, to register the reference scan to the follow-up scan and to resolve the misalignment error between scans. Registration is performed on the lung masks to avoid registration errors that may arise due to the pathological changes in the lung parenchyma. An example of aligned CT scans from different time points are shown in Figure~\ref{fig:regis}.

\begin{figure}[t]
\centering
\makebox[\linewidth][c]{\includegraphics[width=\linewidth]{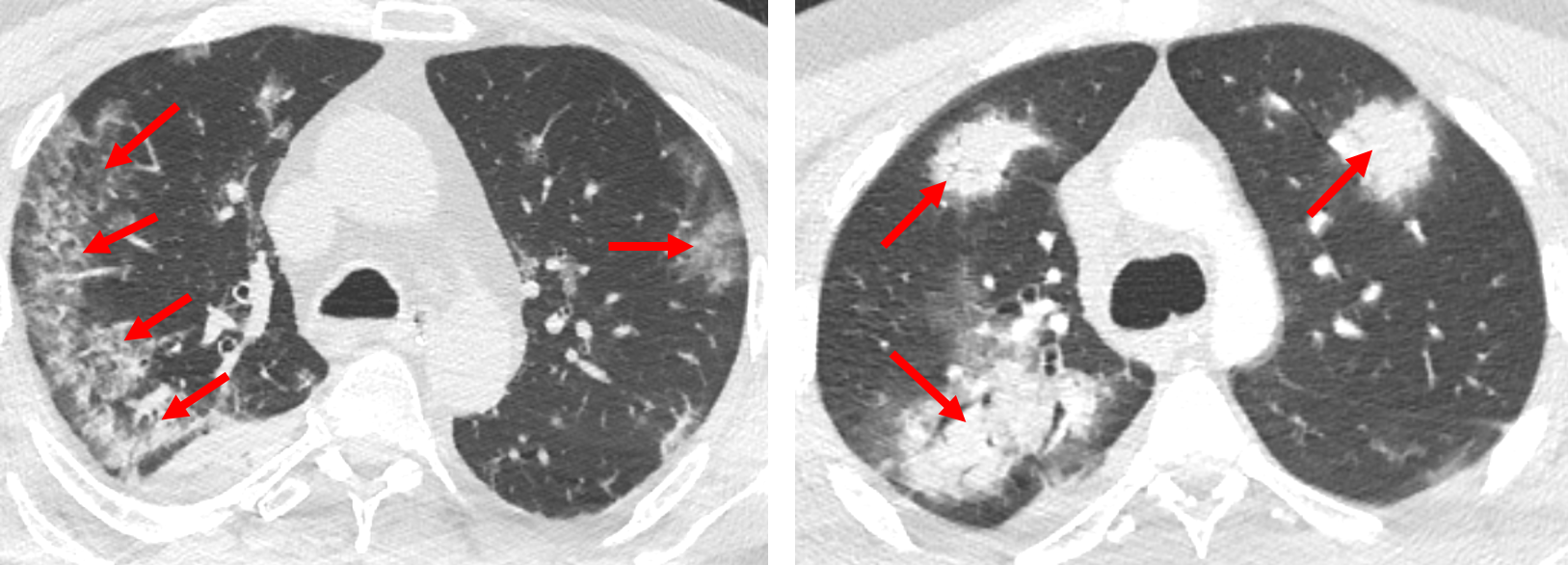}}%
\caption{Examples of GGO and CONS on axial slices of COVID-19 patients' lung CTs. Arrows point to the infection areas. \textit{Left:} GGO \textit{Right:} CONS.}
\label{fig:eg}
\end{figure}

According to \cite{Hefeda2020}, GGO is the most common findings in COVID-19 patients' CT scans, followed by CONS. Figure~\ref{fig:eg} shows examples of GGO and CONS from our dataset. For our experiments, we only segment GGO and CONS, due to the low occurrences of PLEFF in the patient cohort of the dataset. 

For the 30 patients, the reference and follow-up CT scans of each patient after registration have an average structural similarity index (SSIM)~\cite{wang2004image} of 29.71\%. This shows that the scans taken at different time points broadly differ from one another perceptually. Besides that, the average change in the percentage of GGO and CONS in the patients' lung CTs from different time points is 13.68\% and 6.59\%, respectively. This indicates the noticeable difference in the disease progression over time in the dataset. Table \ref{data_1} shows the percentage of GGO and CONS in the lungs of the patients at each timestep.

\begin{table}[t]
\caption{Percentage of GGO and CONS in the lungs of patients. $T$ refers to the timestep of the scans. $n$ is the number of patient volumes at each timestep}
\vspace{1mm}
\label{data_1}
\resizebox{\linewidth}{!}{\begin{tabular}{l cc cc cc}
\toprule
Radiomic & Average & Std. Dev. & Average & Std. Dev. & Average & Std. Dev. \\
\addlinespace[2pt]
\hline
\midrule \addlinespace[2pt]
 & \multicolumn{2}{c}{T-1 (n=30)} & \multicolumn{2}{c}{T-2 (n=30)} & \multicolumn{2}{c}{T-3 (n=5)} \\  
\cmidrule[0.4pt](lr{0.125em}){2-3}%
\cmidrule[0.4pt](lr{0.125em}){4-5}%
\cmidrule[0.4pt](lr{0.125em}){6-7}%
GGO & 15.23\% & 14.76\% & 20.17\% & 18.07\% & 15.02\% & 11.91\% \\ \midrule \addlinespace[2pt]
 & \multicolumn{2}{c}{T-1 (n=30)} & \multicolumn{2}{c}{T-2 (n=30)} & \multicolumn{2}{c}{T-3 (n=5)} \\ 
\cmidrule[0.4pt](lr{0.125em}){2-3}%
\cmidrule[0.4pt](lr{0.125em}){4-5}%
\cmidrule[0.4pt](lr{0.125em}){6-7}%
CONS & 6.52\% & 7.07\% & 8.15\% & 11.51\% & 11.28\% & 11.07\% \\ \bottomrule 
\end{tabular}}
\end{table}

The training set is made up of 16 patients (37 volumes), with training (n=12) validation (n=4) split. Our model is tested on an independent test set consisting of 14 patients (28 volumes). 

\subsection{Experimental Settings}
\subsubsection{Evaluation Metrics}
The segmentation performance of the models is evaluated using the following metrics. 
\begin{itemize}
    \item \textbf{Dice Similarity Coefficient (DSC)} is a statistical measure of the similarity between two segmentations.
    \begin{equation}DSC = \frac{2\text{TP}}{\text{2TP + FP + FN} } \end{equation} 
    \item \textbf{Positive Predictive Value (PPV, or precision)} displays the fraction of correctly segmented regions over all predicted segmentations.
    \begin{equation}PPV = \frac{\text{TP}}{\text{TP + FP} }\label{eqn:pre}\end{equation}
    \item \textbf{True Positive Rate (TPR, recall or sensitivity)} shows the proportion of correct segmentation outputs with respect to the ground truth.
    \begin{equation}TPR = \frac{\text{TP}}{\text{TP + FN} }\label{eqn:rec}\end{equation}
    \item \textbf{Volume Difference (VD)} is calculated as the absolute difference in the predicted lesion segmentation volume and ground truth lesion segmentation volume over the ground truth lesion segmentation volume.
    \begin{equation}VD = 100 \times {\frac{\left |\text{Lesion\_volume$_{pred}$ - Lesion\_volume$_{gt}$}\right |}{\text{Lesion\_volume$_{gt}$} }}\label{eqn:vd}\end{equation}
\end{itemize}

\subsection{Ablation Study}
In order to study how the additional information concatenated to the inputs influences the model's segmentation performance, an ablation study was carried out using our longitudinal COVID-19 dataset. In addition to the longitudinal baseline network~\cite{Denner2020}, we also implemented a static version of the network for comparison. long\_edit\texttt{+}ref\_seg is the baseline long.\texttt{+}ref\_seg extended for interactive segmentation. The proposed model additionally incorporates past prediction outputs of the target as additional information to guide the following segmentation, and static\_edit is the version of the proposed model without reference information. Table~\ref{tab0} summarizes all tested models and their inputs. The reference manual segmentation in the table refers to ground truth masks of the reference images, whereas the edit masks contain the user feedback on the target segmentation. 

\begin{table}[t]
  \begin{center}
    \caption{Input for different models tested in ablation study}\label{tab0}
    \vspace{1mm}
    \resizebox{\linewidth}{!}{\begin{tabular}{lccccc}
    \toprule 
    & \multicolumn{5}{c}{\textbf{Model input}} \\
    \cmidrule{2-6}
    \textbf{Model name} & Target & Reference & Ref. manual  & Edit  & Target previous  \\ 
    & image & image & segmentation & mask & segmentation\\
    \midrule 
    Baseline static network & \checkmark &  & & & \\ 
    Baseline long. network & \checkmark & \checkmark & & & \\ 
    Baseline long.$+$ref\_seg & \checkmark & \checkmark & \checkmark & & \\ 
    static\_edit & \checkmark &  & & \checkmark & \checkmark\\ 
    long\_edit$+$ref\_seg & \checkmark & \checkmark & \checkmark & \checkmark & \\ 
    \textbf{Proposed} & \checkmark & \checkmark & \checkmark & \checkmark & \checkmark \\ 
    \bottomrule 
    \end{tabular}}
  \end{center}
\end{table}

\newpage
As it can be inferred from Table \ref{tab4}, the baseline models performed better than the interactive segmentation models in the first segmentation round without edits and the baseline longitudinal models have higher Dice scores compared to the baseline static model. Among the longitudinal baseline models, concatenating reference segmentation to the input CTs can further improve its GGO Dice by 1.44\% and CONS Dice by 0.76\%. Comparing the interactive segmentation models, the longitudinal interactive segmentation models output better initial segmentation than the static model. The proposed model's GGO Dice is 8.45\% higher than the static model, whereas the long\_edit\texttt{+}ref\_seg model has an improvement of 15.7\% in its CONS Dice over the static model.

\begin{table*}[t]
\caption{Evaluation results on the test set before user interactions. Values displayed are the mean and standard errors. Bold values represent the best results for each metric. Empty masks are used in place of the edit mask and target image previous segmentation mask for static\_edit, long\_edit\texttt{+}ref\_seg and the proposed model.}
\vspace{1mm}
\label{tab4}
\resizebox{\linewidth}{!}{
\centering
\begin{tabular}{l|c|c|c|c|c|c|c|c}
\toprule 
\multirow{2}{2.2cm}{\textbf{Model}}  & \multicolumn{2}{c|}{\textbf{Dice (\%)}}  & \multicolumn{2}{c|}{\textbf{PPV (\%)}} & \multicolumn{2}{c|}{\textbf{TPR (\%)}} & \multicolumn{2}{c}{\textbf{VD (\%)}} \\ 
\cmidrule{2-9}
& \textbf{GGO} & \textbf{CONS} & \textbf{GGO} & \textbf{CONS} & \textbf{GGO} & \textbf{CONS} & \textbf{GGO} & \textbf{CONS} \\ 
\midrule 
\multicolumn{6}{l}{\textbf{Non interactive methods}} \\
\midrule 
Baseline static network & 44.15 $\pm$ 3.33 & 19.75 $\pm$ 4.89 & 62.12 $\pm$ 2.89 & \textbf{42.86} $\pm$ 8.53 & 38.98 $\pm$ 5.09  & 14.96 $\pm$ 3.60 & 49.98 $\pm$ 6.07 & 102.0 $\pm$ 28.84 \\ 
Baseline long. network & 45.42 $\pm$ 3.03 & 27.63 $\pm$ 5.61 & \textbf{67.34} $\pm$ 3.03 & 41.35 $\pm$ 8.01 & 37.96 $\pm$ 4.46 & 23.54 $\pm$ 4.32 & 48.56 $\pm$ 6.09 & 93.18 $\pm$ 38.04\\ 
Baseline long.+ref\_seg & \textbf{46.86} $\pm$ 3.12 & 28.39 $\pm$ 6.39 & 62.61 $\pm$ 3.86 & 42.12 $\pm$ 8.53 & \textbf{41.63} $\pm$ 4.16 &  24.36 $\pm$ 5.61 & \textbf{43.31} $\pm$ 6.53 & 90.36 $\pm$ 39.43\\ 
\midrule 
\multicolumn{6}{l}{\textbf{Interactive methods}} \\
\midrule 
static\_edit & 35.97 $\pm$ 4.20 & 13.99 $\pm$ 3.51 & 66.82 $\pm$ 3.01 & 41.20 $\pm$ 8.45 & 29.03 $\pm$ 4.99 & 9.28 $\pm$ 2.27 & 59.12 $\pm$ 7.20 & 98.11 $\pm$ 22.21\\
long\_edit+ref\_seg & 37.70 $\pm$ 3.89 & \textbf{29.69} $\pm$ 5.86 & 64.66 $\pm$ 3.03 & 37.85 $\pm$ 7.60 & 30.39 $\pm$ 4.46 & \textbf{26.80} $\pm$ 5.35 & 55.46 $\pm$ 6.75 & 64.88 $\pm$ 27.62\\ 
\textbf{Proposed} & 44.42 $\pm$ 3.61 & 23.94 $\pm$ 6.12 & 62.65 $\pm$ 3.49 & 41.89 $\pm$ 8.86 & 40.38 $\pm$ 5.42 & 19.62 $\pm$ 5.92 & 51.84 $\pm$ 7.40 & \textbf{49.71} $\pm$ 7.94\\
\bottomrule 
\end{tabular}
}
\end{table*}

\begin{table*}[t]
\caption{Evaluation results of the interactive methods on the test set with real user interactions. Values displayed are the mean and standard errors. Bold values represent the best results for each metric.}
\vspace{1mm}
\label{tab6}
\resizebox{\linewidth}{!}{
\centering
\begin{tabular}{l|c|c|c|c|c|c|c|c}
\toprule 
\multirow{2}{2.2cm}{\textbf{Model}}  & \multicolumn{2}{c|}{\textbf{Dice (\%)}}  & \multicolumn{2}{c|}{\textbf{PPV (\%)}} & \multicolumn{2}{c|}{\textbf{TPR (\%)}}   & \multicolumn{2}{c}{\textbf{VD (\%)}} \\ 
\cmidrule{2-9}
& \textbf{GGO} & \textbf{CONS} & \textbf{GGO} & \textbf{CONS} & \textbf{GGO} & \textbf{CONS} & \textbf{GGO} & \textbf{CONS} \\ 
\midrule 
\multicolumn{6}{l}{\textbf{Initial segmentation with user edits}} \\
\midrule 
static\_edit & 40.86 $\pm$ 3.16 & 24.33 $\pm$ 3.14 & 67.80 $\pm$ 3.01 & 54.37 $\pm$ 7.45 & 32.77 $\pm$ 4.33 & 17.16 $\pm$ 1.97 & 54.15 $\pm$ 6.31 & 87.99 $\pm$ 20.73 \\
long\_edit+ref\_seg & 44.34 $\pm$ 2.34 & 36.71 $\pm$ 5.57 & 67.03 $\pm$ 3.30 & 44.38 $\pm$ 6.84 & 35.72 $\pm$ 3.39 & 33.91 $\pm$ 5.05 & 48.76 $\pm$ 5.41 & 54.67 $\pm$ 20.78 \\
\textbf{Proposed} & 49.22 $\pm$ 2.58 & 36.33 $\pm$ 5.19 & 64.34 $\pm$ 3.88 & 50.50 $\pm$ 7.10 & 44.64 $\pm$ 4.50 & 30.91 $\pm$ 5.13 & 46.63 $\pm$ 6.54 & 40.08 $\pm$ 6.03 \\
\midrule 
\multicolumn{8}{l}{\textbf{Output segmentation after one round of segmentation refinement by model}} \\ \midrule 
static\_edit & 53.59 $\pm$ 2.33 & 55.48 $\pm$ 4.48 & \textbf{71.63} $\pm$ 4.12 & \textbf{71.34} $\pm$ 4.34 & 44.62 $\pm$ 2.98 & 47.01 $\pm$ 4.68 & 40.21 $\pm$ 3.76 & \textbf{33.68} $\pm$ 5.66 \\
long\_edit+ref\_seg & 54.79 $\pm$ 1.95 & 54.01 $\pm$ 5.03 & 66.70 $\pm$ 3.62 & 51.14 $\pm$ 5.16 & 48.54 $\pm$ 2.54 & \textbf{64.44} $\pm$ 5.67 & \textbf{34.74} $\pm$ 4.55 & 74.43 $\pm$ 36.10 \\
\textbf{Proposed} & \textbf{59.86} $\pm$ 2.57 & \textbf{58.81} $\pm$ 4.36 & 62.06 $\pm$ 4.48 & 58.18 $\pm$ 5.15 & \textbf{61.86} $\pm$ 3.00 & 62.39 $\pm$ 4.06 & 36.39 $\pm$ 11.28 & 34.11 $\pm$ 12.75 \\ 
\bottomrule 
\end{tabular}
}
\end{table*}

The evaluations results in Table \ref{tab4} further revealed that the CONS TPR for the static models are considerably lower than the longitudinal models, with the non interactive methods having a difference of 8.58\% between the baseline static network and baseline long. network. For the interactive methods, the static\_edit model CONS TPR is 10.34\% less than the longitudinal interactive model with lower TPR. These results proves that longitudinal models can improve segmentation of a more complicated class such as CONS.

\subsection{Segmentation Results and Discussions} 
The following evaluations are carried out using the GUI to obtain real user feedback on the segmentation output. 

\subsubsection{Quantitative Results} 
\noindent The aim of the interactive segmentation model is to assist and reduce the user's workload during the segmentation of new data. Thus, its desired function is to take in rough user interactions and improve the segmentation output. To further determine whether the segmentation refinement model serves this purpose, the initial segmentation that is manually corrected and fed into the segmentation refinement model is compared with the model refined segmentation output. The results are shown in Table~\ref{tab6}.

As can be seen from the results in Table~\ref{tab6}, the interactive segmentation refinement model was able to use the previous segmentation results and user feedback to improve the segmentation of the target scans. Out of the three models, the proposed model has the highest Dice scores for GGO and CONS after one round of segmentation refinement. However, by comparing the change in Dice of the edited initial segmentation by the user and refined segmentation by the model, the static interactive segmentation model showed the most considerable improvement in its segmentation output, with an increase of 12.73\% in GGO Dice and 31.15\% in CONS Dice, whereas the proposed model showed an increase of 10.64\% in GGO Dice and 22.48\% in CONS Dice.

\begin{figure*}[t]
     \centering
     \begin{minipage}[t]{0.47\textwidth}
     {\includegraphics[scale=0.45]{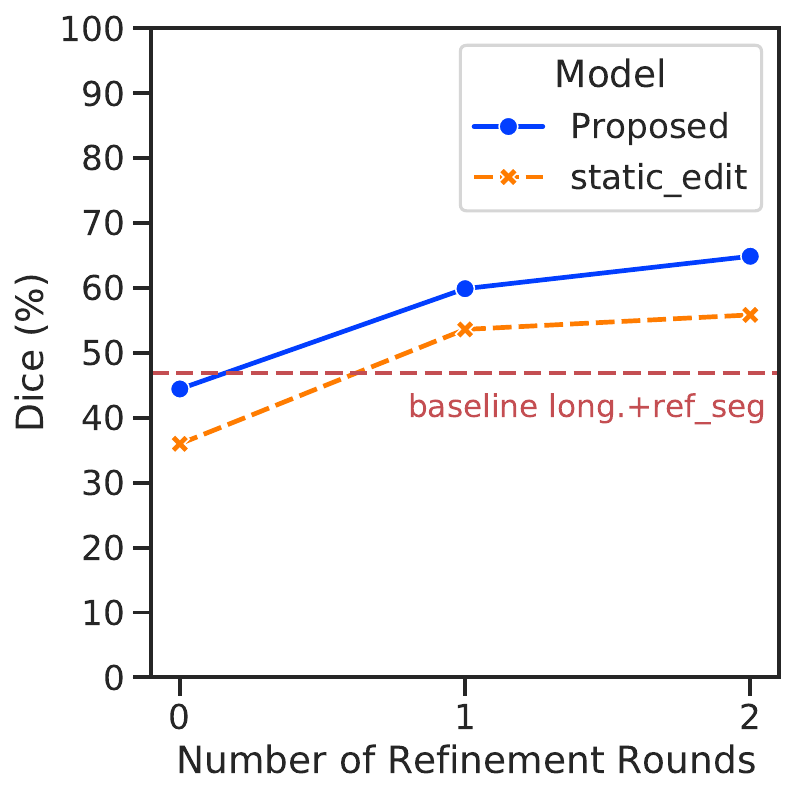}}
     \label{<figure1>}
     \end{minipage} 
     \begin{minipage}[t]{0.47\textwidth}
     {\includegraphics[scale=0.45]{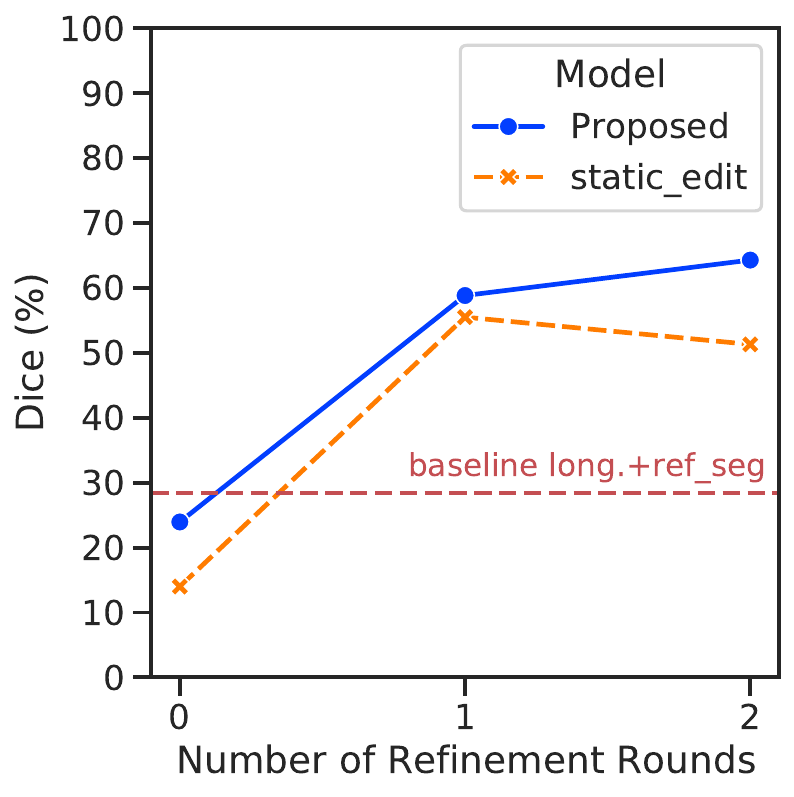}}
     \label{<figure2>}
     \end{minipage} 
     \caption{Average change in Dice vs Number of Refinement Rounds for 14 test patients. \textit{Left:} class GGO and \textit{Right:} class CONS.}
     \label{fig:39}
\end{figure*}

The proposed model and static$\_$edit model are further evaluated with another round of segmentation refinement. Figure \ref{fig:39} shows how the Dice of 14 patients change after each segmentation refinement rounds with real user edits. The best baseline model: baseline long.$+$ref\_seg is added for comparison.

Experimental results showed that the proposed model improves the Dice by a significant amount just after two rounds of segmentation refinement with real user feedback. The total average increase of the segmentation Dice after two rounds of segmentation refinement is 20.44\% for GGO and 40.33\% for CONS, with an average increase of 4.99\% for GGO and 5.46\% for CONS between the first and second segmentation refinement rounds. As for the static\_edit model, after two rounds of segmentation refinement, the GGO and CONS Dice is 9.03\% and 12.98\% lower than the proposed model.

From the plots in Figure~\ref{fig:39}, it is visible that the proposed model is superior in terms of segmentation refinement performance compared to the static interactive segmentation model. The static\_edit model is shown to be unable to further refine the segmentation correctly at the second refinement step; the GGO segmentation Dice improved, whereas the CONS segmentation Dice dropped. Closer inspection of the segmentation output shows that the static\_edit model is more likely to wrongly classify regions, which leads to the Dice decrease. As for the proposed model, the inclusion of reference scan information improved its segmentation classification.  

Since the degree of infection severity varies among the test patients, which influences the output segmentation, the amount of rough user edit strokes needed to edit a patient's initial segmentation range from 1 to 10 per edited slice for the proposed model. 
In most cases, the increase in Dice is larger for CONS after refinement with the model. Qualitative results of the initial segmentation showed that CONS is frequently under segmented due to its high similarity to the background class (blood vessels and walls of airways have a very similar Hounsfield unit histogram) making it difficult for the model to segment the region correctly without user guidance. 
\\

\subsubsection{Qualitative Results}
\begin{figure}[t]
\makebox[\linewidth][c]{\includegraphics[width=\linewidth]{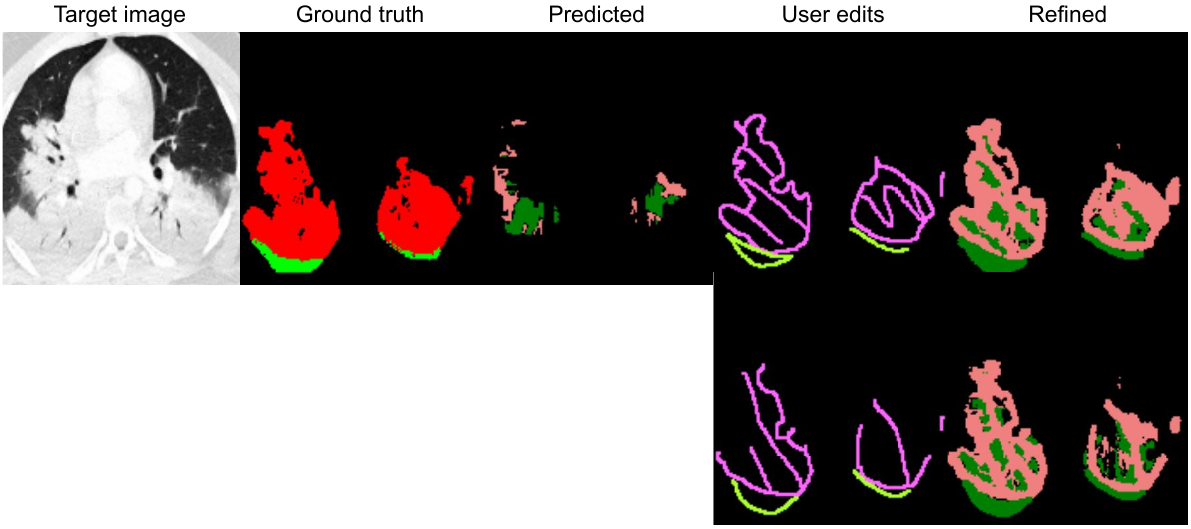}}
\caption{Example of how the refined segmentation differs when different types of user edits (top and lower rows) are drawn to refine the segmentation. (\textit{Red:} GGO ground truth, \textit{Green:} CONS ground truth, \textit{Pink:} predicted GGO, \textit{Dark green:} predicted CONS, \textit{Magneta:} foreground edit for GGO, \textit{Neon green:} foreground edit for CONS)}
\label{fig:34}
\end{figure}

\begin{figure*}[t]
\centering
\includegraphics[width=\textwidth]{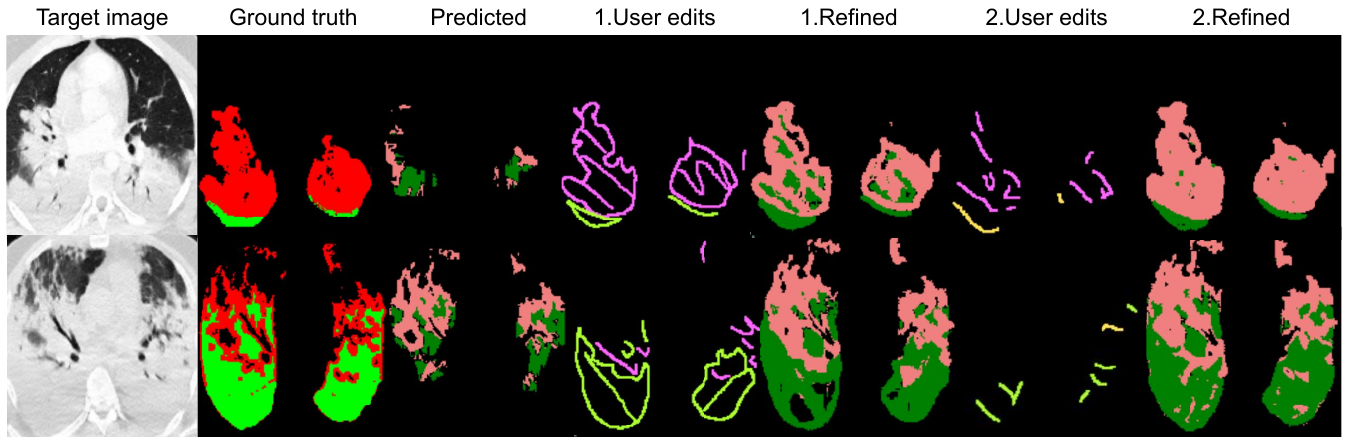}
\caption{Examples of two rounds of segmentation refinement, resulting in increasingly better results.}
\label{fig:35}
\end{figure*}

Qualitative results from the proposed model are presented in this section.
\\
\noindent\textbf{Comparison of Edits} \\
Figure~\ref{fig:34} shows the different segmentation refinement output given different types of user edits. As can be observed in the lower segmentation output, it is sufficient to draw the outer outline of the infected regions to separate them from the background. The model is able to further segment the unedited areas, but it is often incorrectly classified. However, as shown in Figure \ref{fig:35}, the wrong segmentation can be corrected in the following refinement round.

A possible explanation for the wrong segmentation of GGO as CONS is due to their feature similarity in the CT scans. One of the main characteristics that differentiate GGO from CONS is the location of the segmentation. CONS is often located at the bottom of the axial lung CT. But in severe cases, it can be found higher up and its borders are often GGO as shown in the second row of Figure \ref{fig:35}.
\\

\begin{figure*}[t]
\makebox[\textwidth][c]{\includegraphics[width=\textwidth]{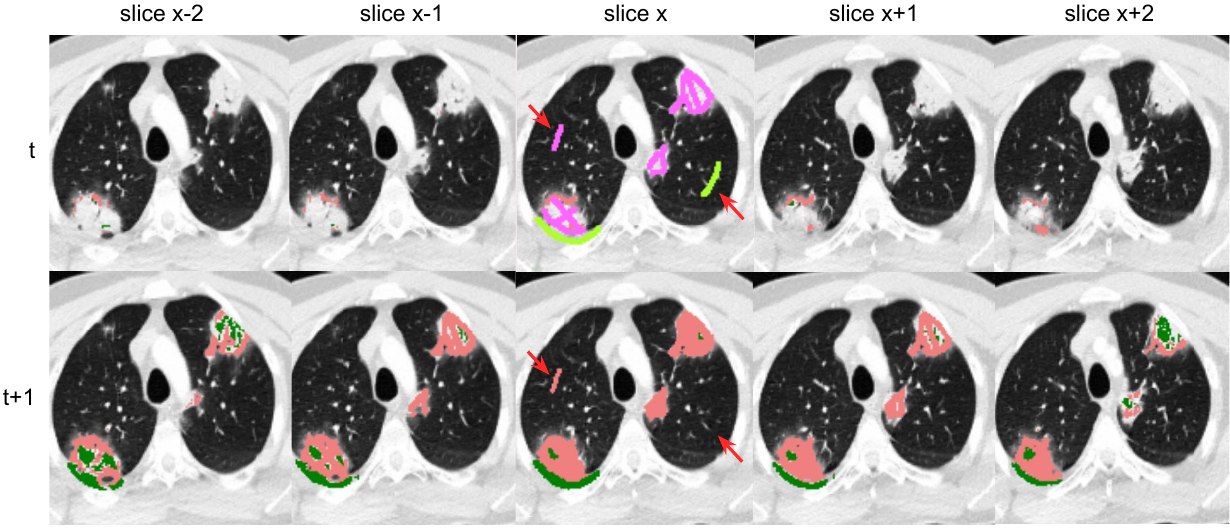}}
\caption{Example of how edits on one slice are automatically propagated to other slices. After the initial prediction, edits are drawn on one slice. Red arrows point to false edit locations. Interestingly, the model learns that false edits should not be propagated further than one slice by the model.}
\label{fig:37}
\end{figure*}

\newpage
\noindent\textbf{Robustness Testing} \\
A robustness test is carried out to evaluate how well the 2.5D model that uses stacked slices from three anatomical views as input is able to propagate edits on one slice to another.

Figure~\ref{fig:37} displays the limited automatic propagation of edits from one slice to other slices. The further the slices are from the edited slice, the more segmentation are wrongly classified even though the correct regions are segmented and some regions are undersegmented. However, false edits are not propagated further than one slice away by the model. This shows the model's ability to detect edits that are incorrect by considering the features of the CT scans.  
\\

\begin{figure}[t]
\makebox[\linewidth][c]{\includegraphics[width=\linewidth]{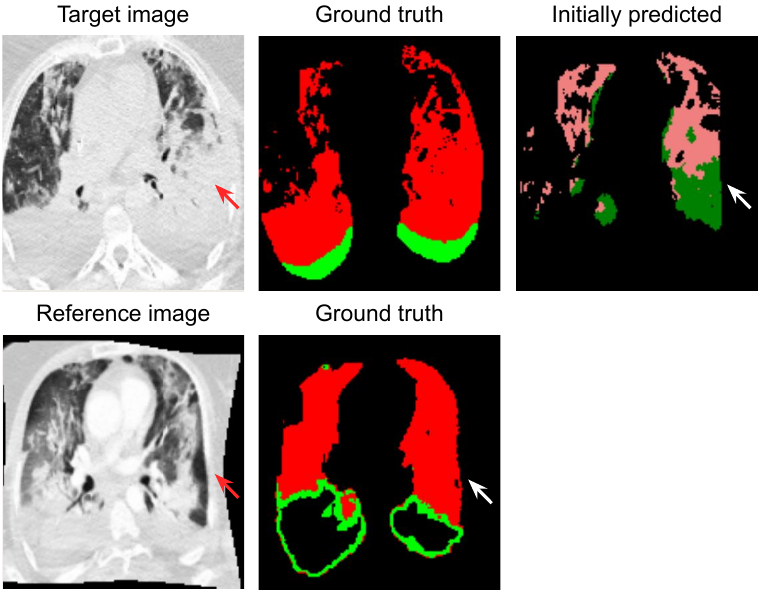}}
\caption{Example of how the target's initial prediction is influenced by the reference image in a case where there is a misalignment between the reference and target image. The red arrows point to regions where the CT scans differ the most. The white arrows point to the interested regions.}
\label{fig:32}
\end{figure}

\noindent\textbf{Influence of Reference Data on Target Segmentation} \\
In some cases where a patient's reference scan is not correctly deformed to align with the follow-up scan, the initial segmentation of the follow-up scan for more difficult regions can be negatively affected by the reference CT. Figure~\ref{fig:32} shows some abnormalities in the initially predicted segmentation of the target image. The infected lower part of the target CT that is harder to discern from the background class is segmented according to the reference image. As observed in the target's initial predicted segmentation, its left outline looks similar to the reference image.

\begin{figure}[t]
\makebox[\linewidth][c]{\includegraphics[width=\linewidth]{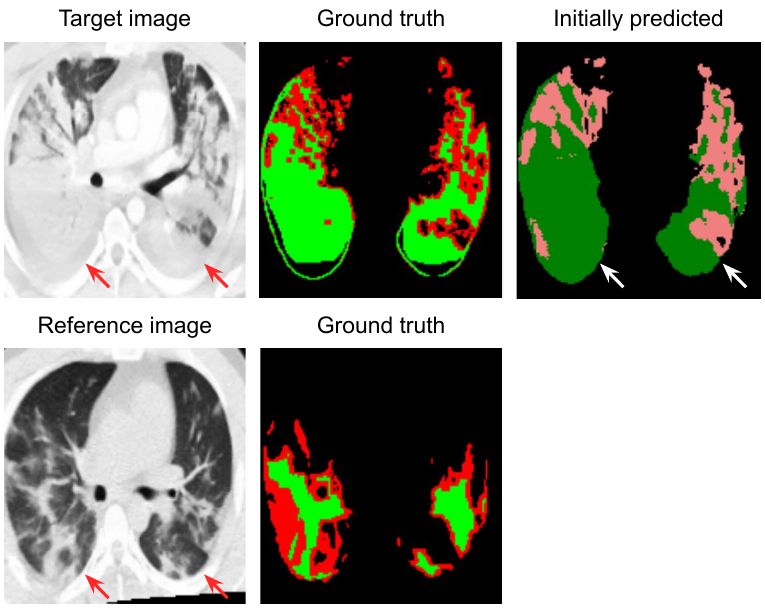}}
\caption{Example of a case where the reference image and target image are correctly aligned. The model segmented the difficult areas better compared to Figure \ref{fig:32}. The red arrows point to regions where the CT scans differ the most. The white arrows point to the interested regions.}
\label{fig:40}
\end{figure}

In another example, when the reference image is correctly aligned with the target image, as shown in Figure~\ref{fig:40}, the reference image can serve as a guide for the model to segment regions that are difficult for the model as previously mentioned, such as areas that look similar to the background class. In this example, the model produces a more accurate initial segmentation of the target scan. 

\subsection{Limitations and Future Work}
One of the limitations that we faced is the availability of a larger longitudinal COVID-19 dataset. Also, label noise is probably present in our data, since for severe cases, even expert radiologists struggle to separate in particular CONS from the background class and PLEFF. In future work, the potential of this method on improving segmentation results of other longitudinal medical images and in different clinical contexts can be further explored. The problem of wrongly classified segmentation in multiclass segmentation can be additionally examined through using an ensemble of binary interactive segmentation models for each foreground class. Furthermore, it would be interesting to test the proposed method with other state of the art model architectures. Besides, this paper also reveals limitation of a 2.5D model in propagating edits from one slice to other slices further away, a 3D implementation of the interactive segmentation method which is not in the scope of this paper would perhaps be able to counter this issue at the cost of requiring bigger training datasets.

\section{Conclusion}
In this paper, an interactive segmentation method with 2.5D longitudinal network is proposed. Through concatenating the previous time-point reference segmentation mask, segmentation output of target image from past segmentation rounds and user interactions mask to target and reference images, all past information is fully utilized as input data to the model to improve segmentation results. Experiments on our in-house longitudinal COVID-19 dataset show that a large improvement in the Dice of both classes is obtained after one round of interactive segmentation refinement. Besides that, the proposed longitudinal interactive segmentation refinement model's segmentation performance is superior compared to the static version of the interactive model. This concludes that, with the availability of longitudinal data, existing segmentation models can be easily adapted through our method and trained end-to-end for interactive segmentation refinement. 

\bibliography{egbib}

\end{document}